\pdfoutput=1 % only if pdf/png/jpg images are used
\documentclass{JINST}

\title{The Square Kilometer Array: cosmology, pulsars and
other physics with the SKA}

\author{F. Combes$^a$\\
\llap{$^a$}Observatoire de Paris, LERMA, and College de France,\\
 61 Av. de l'Observatoire, 75014 Paris, France\\
E-mail: \email{francoise.combes@obspm.fr}}

\abstract{SKA is a new technology radio-telescope array, about two orders
of magnitude more sensitive and rapid in sky surveys than
present instruments.  It will probe the dark
age of the universe, just afer recombination, and during the epoch of
reionisation (z=6-15); it will be the unique instrument to map the atomic
gas in high redshift galaxies, and determine the amount and distribution
of dark matter in the early universe.  Not only it will detect and measure the redshifts
of billions of galaxies up to z=2, but also it will discover and monitor around 20 000
pulsars in our Milky Way. The timing of pulsars will trace the stretching of space,
able to detect gravitational waves. Binary pulsars will help to test gravity in strong fields,
and probe general relativity. 
These exciting perspectives will become real beyond 2020.}


\begin{document}

\section{Main questions in Cosmology}\label{sec:intro}

We know now with great precision the content of the universe: with 70\% dark
energy, and 25\% dark matter,  the dark sector is dominating by far, while the
baryons are only 5\% of the total. However, the nature of these components 
is not yet known, and one of the main goals of future instruments is
to determine their evolution in time, i.e. with redshift, to constrain their nature.
 Is the dark energy varying with time, ot can it be reduced to a cosmological 
constant? Can we compare visible and dark matter at any redshift?
  Several missions, such as Euclid, or LSST, will gather data on billions of galaxies to 
 trace with precision the evolution of the dark sector. SKA can bring complementary
data in tracing as many galaxies at other wavelengths.

How is the Universe re-ionized? By mapping the neutral gas in its 21cm fundamental line,
redshifted in the meter range, SKA will be unique to answer with high sensitivity 
and spatial resolution. The observations will cover the 
end of the dark age: cosmic dawn with line absorption, and Epoch of Reionization (EoR)
with line emission.

How do baryons assemble into the large-scale structures?
The observations of the atomic gas reservoirs at high redshift, 
which are not yet available, will considerably enlight galaxy formation 
and evolution, the relative role of mergers or cold gas accretion in mass assembly,
the star formation history and quenching, the role of 
environment in galaxy groups and clusters.

Black hole growth is accompanying closely bulge growth through star formation.
It is likely that black holes and bulges compete for feeding, but also 
that implied nuclear activity (AGN) moderate or quench star formation
through outflows, which will be studied at all redshifts.

Strong-gravity will be probed around pulsars and black holes. The discovery
of thousands of new pulsars and their timing will provide a
gravitational wave telescope.

\section{How to probe the content and evolution of the universe}\label{sec:DE-BAO}

Looking far away with sensitive telescopes, it has been possible
to look back in time, almost through the 13.8 billion years of its existence.
 The succession of events is now well known: following the hot Big-Bang,
the plasma recombines 370 000 yrs after, and then begins the dark age, when 
there is no star yet to shine. Progressively, the baryons fall into galaxies and 
the gas collapses to form stars, whose UV light re-ionizes 
patches of universe around them. The epoch of re-ionization is
expected to last between 0.5 and 1 billion years. Although active nuclei powered by 
super-massive black holes already exist and provide ionizing photons, 
the most likely agents of the re-ionization of the universe are stars in galaxies.
After this cosmic renaissance, galaxies evolve and assemble mass, 
through accretion and hierarchical merging. 
Large scale structures up to galaxy clusters progressively emerge,
and their precise growth rate will probe possible gravity laws modified
beyond general relativity.

What is the fate of our universe? The standard candles SNIa have
revealed the acceleration of expansion.  Some of the best results
on these tracers come from the
2003-2008 SNLS survey, a French-Canadian collaboration
on the CFH telescope (about 500 SNIa, 
Sullivan et al. 2011, Conley et al. 2011, Betoule et al. 2014).
Assuming a flat universe, it was possible to constrain values
of the matter density $\Omega_m$ and the parameter $w$ from
the equation of state of the dark energy: P=$w \rho$. The value
$w=-1$ compatible with a cosmological constant is still one of the most likely,
within a 6\% error bar.

When all the indicators are taken into account together,
the supernovae Ia, the cosmic microwave background (CMB),
the weak lensing and the baryonic acoustic oscillations for the
large-scale structures, the concordance model begins
to constrain more parameters of the dark energy 
equation of state  P=$w \rho$. The $w$ parameter
is developped in $w(a)= w0 +wa (1-a)$,
where $a$ is the expansion radius, normalised to 1 at z=0.
Before Planck, Kowalski et al. (2008) found a weak evolution
with redshift possible, while there is no 
evidence for a dynamical dark energy with Planck data (2014). 

However, the CMB gives only an instantaneous view of the universe,
370 000 yrs after the Big Bang. The anisotropies measured in the CMB,
coming from acoustic oscillations of photons and baryons together,
provide a tremendous amount of data, on the curvature, the density
of baryons, dark matter, and they already reveal some departures from
the standard model predictions. For instance there are missing large-scale
structures: there is not enough power at low spatial frequency (low-$l$)
in the power-spectrum of the CMB signal.

To determine the evolution of dark energy, and therefore its nature,
it is essential to measure the rate of expansion of the universe at any epoch.
This can be done with the detection of baryonic acoustic oscillations (BAO), at all redshift.
 The maximum size of these oscillations is well known, it is the sound horizon
at the epoch of recombination, at t=370 000 yrs. The sound had time to run 150 Mpc,
and the measurement of this size, then frozen in the comoving volume,  
in the power-spectrum of galaxies at any z will serve as a ruler to estimate
the expansion rate.

\subsection{BAO standard ruler}

The first detection of BAO was done in the local universe by Eisenstein et al. (2005)
with the Sloan spectroscopic  survey, and in particular the luminous red galaxies (LRG).
The region covered included about 47 000 galaxies, with redshifts between 0.2 and 0.5. 

Already Alcock \& Paczynski (1979) had proposed a simple way
to probe the value of the cosmological constant, when galaxies 
are expanding in a ring structure: it is possible to compare the radius of the ring
in the line of sight direction c$\Delta z$/H, and in the plane of the sky 
$\Delta \theta$ D (where D is the angular distance). Since we are measuring 
the baryonic structures over the dynamics of the total matter, it will be possible
to reach also the bias, i.e. to know how much the baryons trace or not the dark matter. 

Today, there are many ongoing or future BAO surveys, like the Sloan BOSS,
the DES or BigBOSS survey, Euclid, LSST projects and efficient reduction and reconstruction 
methods  (e.g. Burden et al. 2014). All the surveys will be carried out in optical
or infrared wavelengths. It is important that billions of galaxies could also
be traced in a complementary way, with the atomic gas and the 21cm line. 
The biases will be different, for instance, HI-rich galaxies tend to avoid galaxy clusters,
while the infrared surveys on the contrary are biased towards high surface-densities of
galaxies.

There are several ways that SKA can contribute significantly to the
BAO surveys. The first phase SKA-1 can cover three types of
surveys, differing by their depth: an
all sky survey (3$\pi$ sr), providing  4 10$^6$  galaxies at  z$\sim$ 0.2 ,
a wide-field survey (5000 deg$^2$) with  2 10$^6$  galaxies
at z$\sim$0.6, and a
deep-field survey (50 deg$^2$) with  4 10$^5$  galaxies at z$\sim$0.8.
Alternatively, an intensity mapping can be carried out over 25 000 deg$^2$,
and will bring a competitive constraint on the expansion rate at z$\sim$ 2 (Bull et al. 2015).
The second phase SKA-2 will be able to provide a redshift survey over 30 000 deg$^2$,
yielding much better results than other experiments, for redshifts up to 1.4.

The large galaxy redshift surveys foreseen with SKA will not only be useful
for BAO, but also for other tools, like the Redshift Space Distortions (RSD),
tracing peculiar velocities off the Hubble flow, either random (finger-of-god effect in galaxy clusters),
or more coherent (infall of galaxies on clusters, Kaiser effect). Cross-correlation with the optical
surveys, such as Euclid, will enrich the information (Abdalla et al. 2015).

About the weak shear tool, SKA can provide high-resolution continuum
images of 10 billion galaxies.

\begin{figure}[tbp] % figures (and tables) should go top or bottom of
                    % the page where they are first cited or in
                    % subsequent pages
\centering
\includegraphics[width=1.0\textwidth]{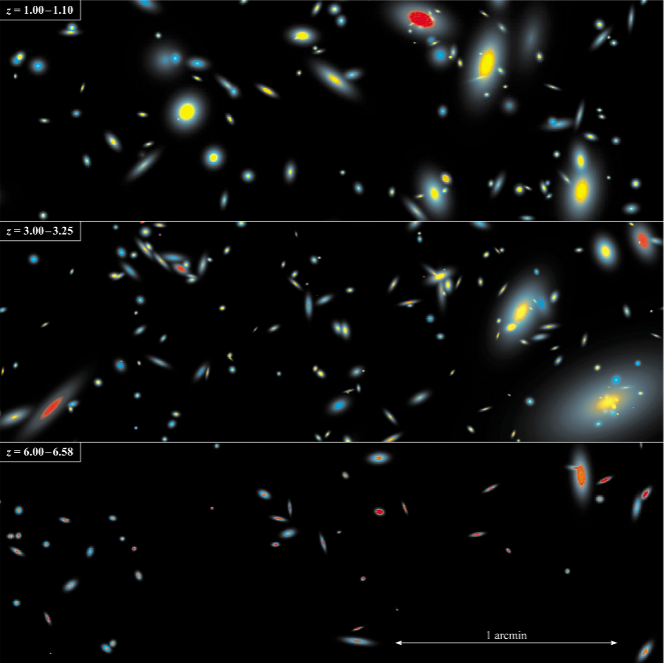}
\caption{ Simulated sky over a field of 3 x 1 arcmin$^2$. From top to bottom
are represented 3 different redshift ranges (z=1-1.10, 3-3.25 and 6-6.58), with depth corresponding
all to 240 Mpc. Blue colors indicate HI-21 cm line flux, and the orange-red colours to the CO line
(molecular gas).
From Obreschkow et al.  (2009b).}
\label{fig:virtual}
\end{figure}

\subsection{Continuum surveys with SKA-1}

In 2yrs of all-sky survey, the goal with SKA-1 is to 
achieve 2 $\mu$Jy rms, which will allow to
detect at more than 10$\sigma$ $\sim$4 galaxies per arcmin$^2$. 
 The spatial resolution and the quality of the synthetized beam,
will allow to measure the weak lensing, and the Integrated Sachs Wolfe
(ISW) effect.
With almost uniform sky coverage of 3$\pi$ sr, a 
total of 0.5  billion radio sources will be detected.
In the
wide-field (5000 deg$^2$) survey, $\sim$6 galaxies per arcmin$^2$
can be reached above  10$\sigma$ with a  2 $\mu$Jy rms, while in
the deep-field (50deg$^2$) with 0.1 $\mu$Jy rms, this
leads to $\sim$20 galaxies arcmin$^2$. 

These numbers can be compared to the
present status of radio surveys.
In the Hubble deep field north, an area 
of 5 x 5 arcmin has been observed during 50 hours,
with the VLA by Fomalont et al. (1997); 
6 sources were detected with  a flux at 8.4GHz larger than > 12 $\mu$Jy.

Radio-galaxies are of different types, the brightest are AGN (active nuclei),
but they can also be ordinary star forming galaxies or starbursts. These large
surveys whill reveal how they trace the bulk of the mass at any redshift. In any 
case, the bias of these sources, the way they trace the dark matter
distribution, will be quite different than for optical/infrared surveys
(Jarvis et al. 2015). 
Brown et al. (2015) have shown how SKA will bring new perspectives
to weak-lensing surveys, using polarization and rotation velocity to disentangle
intrinsic alignments, and extend studies at high redshift.

Simulations of the sky at radio wavelengths have been computed by
Jackson (2004), Obreschkow et al. (2009), at different redshifts, and
including the main lines of HI and CO (see figure~\ref{fig:virtual}).

\subsection{Radio relics, radio halos}

\begin{figure}[tbp] % figures (and tables) should go top or bottom of
                    % the page where they are first cited or in
                    % subsequent pages
\centering
\includegraphics[width=1.0\textwidth]{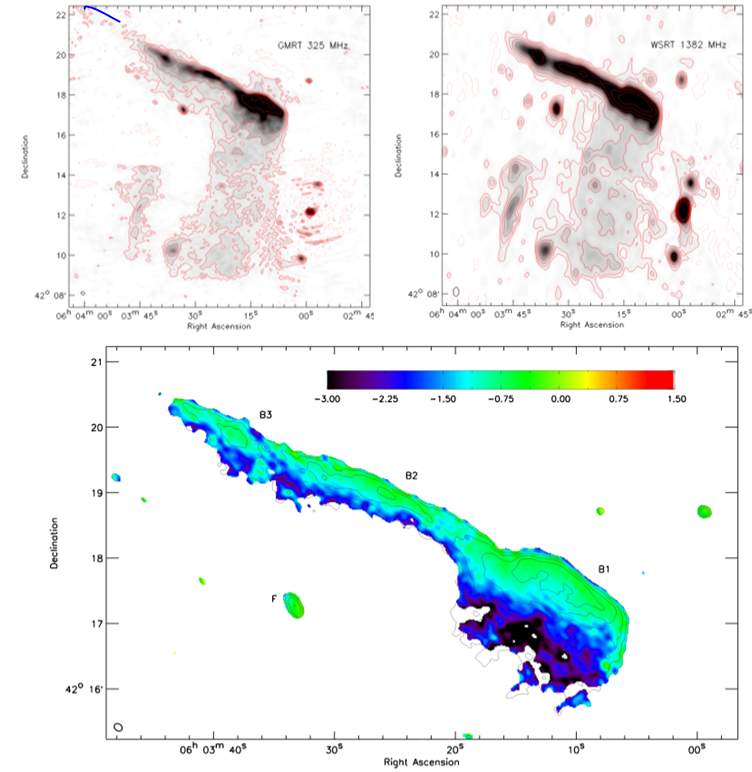}
\caption{ {\bf Top} At left is the GMRT 325 MHz map of the merging cluster 
1RXS J0603+4214, at z=0.225. At right, the 1382 MHz mapt, taken with WSRT.
{\bf Bottom} The colors reproduce the
index of the emission from GMRT between 610 and 325 MHz
(see the color palette at the top), and the contours are from the GMRT 325 MHz image.
Adapted from van Weeren et al.  (2012).}
\label{fig:tooth}
\end{figure}

Radio emission at Mpc scales are often observed in galaxy clusters, and 
are a precious information on the cluster physics, relaxation state and 
magnetic fields. Radio relics are formed as shocks during a cluster merger,
to form a more massive galaxy cluster (e.g. Van Weeren et al. 2012). 
They are elongated and irregular structures, such as in 
 Figure~\ref{fig:tooth}, which shows a radio relic in the form of a tooth
brush. 
In a shock, charged particles are accelerated to relativistic energies.
In the presence of magnetic fields, they radiate synchrotron radiation.
The observations reveal large-scale alignment of magnetic field,
and a strong spectral index gradient.

Radio halos are also diffuse radio emission extending at Mpc scale, 
associated to cluster mergers, but
unpolarised, and located at the center of the clusters. They have a more regular morphology
than radio relics which are often found at the borders of clusters, and are highly polarised.

Giovannini et al. (2015) explore how SKA-1 will be able to detect diffuse radio emission
in low density regions, tracing intergalactic filaments, in galaxy groups and clusters, and 
tackle the evolution of cosmological magnetic fields.

\section{Epoch of Reionization}

The SKA is unique to observe the HI-21cm fundamental line 
redshifted at very high redshift, up to z=20-30. 
 This means that the formation of the first stars and galaxies will
be explored, through the intergalactic medium, which is mostly neutral
atomic hydrogen, until the end of re-ionization at z$\sim$6. 
The UV light from the first stars are ionizing their surroundings, and
numerical simulations have shown how the various ionized patches inflate,
and lead to a progressive percolation of ionized zones (e.g. Iliev et al. 2006,
Baek et al. 2009). 

\begin{figure}[tbp] % figures (and tables) should go top or bottom of
                    % the page where they are first cited or in
                    % subsequent pages
\centering
\includegraphics[width=1.0\textwidth]{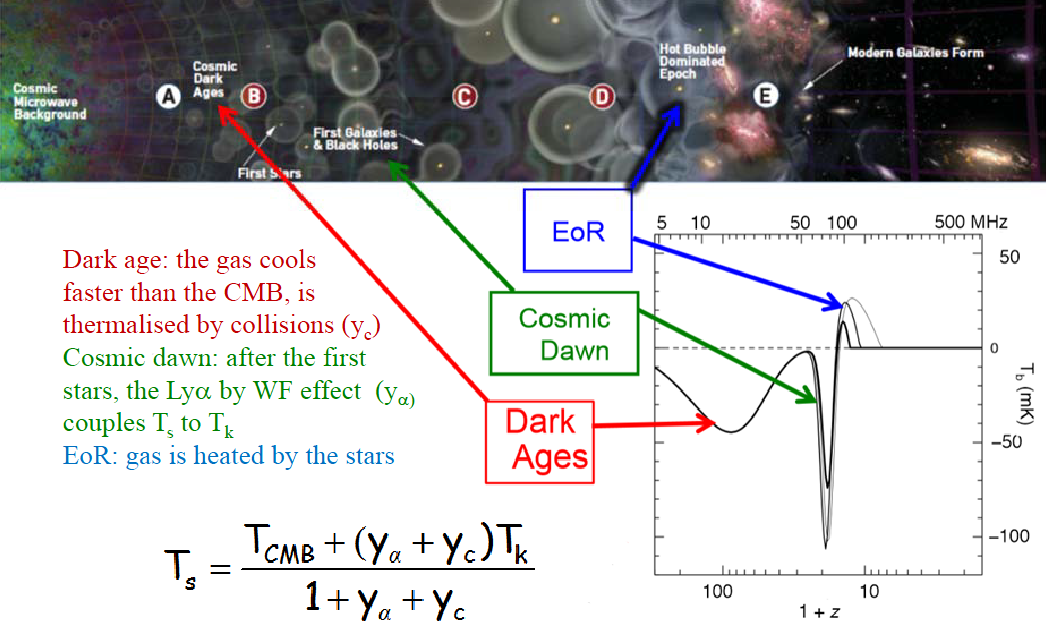}
\caption{Succession of events, from the Big-bang at left, to the EoR (Epoch of Reionization) at right,
during the first billion years of the Universe. In the first phase, the dark ages, 
collisions are still strong enough to couple
the spin (excitation) temperature of the H-atoms to the kinetic temperature of the gas, which 
decreases in $(1+z)^{2}$, faster than T$_{CMB}$ in  $(1+z)$. The expansion of the universe
reduces the collision strength, and the H-atoms spin temperature progressively aligns with
T$_{CMB}$. Then the first stars light up, and emit Lyman$\alpha$ photons, which couple the H-atoms
 to their kinetic temperature again, through the Wouthuysen-Field effect (WF). This is the cosmic dawn, 
and the HI-21cm signal can be detected in absorption. When stars are progressively more 
numerous, the gas is in  majority heated by their radiation, and the 21cm signal can be 
detected in emission.  Adapted from Pritchard \& Loeb (2010).}
\label{fig:EoR}
\end{figure}

Since the gas kinetic temperature falls down in (1+z)$^2$, while the CMB temperature varies as (1+z),
the gas is first colder than the radiation. However the excitation temperature (T$_{spin}$) is equal to the kinetic
temperature only when the medium is dense enough (dark age), or pumped by Lyman-$\alpha$ photons
(cosmic dawn). Without efficient excitation, T$_{spin}$ =T$_{CMB}$, and it is impossible to see any
absorption or emission signal (see figure~\ref{fig:EoR}).  Later on, the gas is heated by the star formation,
and the signal will be seen in emission in the EoR.

Up to now, the EoR signal has not been detected. 
Only simulations have been able to estimate the expected signal.
Although the predicted signal is
strong enough to be observed by present telescopes, the foreground emissions
are at least 4 orders of magnitude higher, and are not easily disentangled. They 
will remain a challenge for SKA (Chapman et al. 2015).

Mellema et al. (2015) describe how HI tomography with SKA-Low is essential to 
characterize the Cosmic Dawn (CD) and Epoch of Reionization (EoR). Path finder
intruments, such as GMRT, MWA, PAPER and LOFAR are able to provide only 
power-spectrum of the signal, but imaging at a given redshift, or make real cubes of data,
will be the privilege of SKA.

Another way to explore the EoR is to image the surrounding of z$>$6 quasars,
and the ionized patch in their environment (Geil \& Wyithe 2008).
These observations will be complementary to those in the optical
or infrared with  JWST and ELT,
detecting the HII regions around high redshift QSO.
Also absorption studies will focus on the 21cm forest,
in analogy to the Lyman-$\alpha$ one (Ciardi et al. 2015).

Since the observation by Planck of a Thompson optical depth significantly lower
than previously thought, it becomes clear now that the first galaxies from z$\sim$10 until
z$\sim$ 6 are perfectly able to re-ionize the universe (e.g. Bouwens et al. 2015).
 The detailed HI tomography will allow to follow the evolution
and growth of these galaxies.

\section{Galaxy formation and evolution}

HI-21cm mapping in galaxies is a unique tool to determine through the 
rotation curve the dark matter distribution in galaxies. Up to now, this has
been possible only in the local universe, by lack of sensitivity. 
With SKA, this will be possible up to z=5, and the evolution of the 
ratio between baryons and dark matter will be known through cosmic time.
  This will help to solve one of the main problem in galaxy formation 
and evolution: why are there so few baryons in galaxies? In the local universe, 
the baryon fraction maximizes at 20\% of the universal baryon fraction,
in galaxies of the Milky Way type, of total mass of $\sim$ 10$^{12}$ M$_\odot$.
The baryon fraction is even much lower in dwarf and massive galaxies.
When do galaxies lose their baryons? or were the baryons never accreted
in galaxy haloes? 

In addition, HI is the fundamental reservoir for star formation and black hole
growth, before it is transformed in the moleuclar phase, at high density in 
the galaxy centers (Leroy et al. 2013). The observation of the neutral gas with high 
resolution up to z=2, corresponding to the peak of activity in the 
star formation history, will help to answer fundamental questions such as:
How galaxies assemble their mass?
How much mass assembled in mergers?
How much through gas accretion and secular evolution?

In the recent years, cold gas accretion from cosmic filaments
have been promoted as the main motor of evolution. Most stars
in the universe have been formed while galaxies are in the main
sequence, and only of the order of 10\% in a starburst mode.
However, the quenching mechanisms have not yet been identified,
although star formation and AGN feedback are suspected.
 The SKA mapping of the morphology and kinematics of the neutral
gas will identify inflows and outflows of gas, and thie evolution
with redshift (Blyth et al. 2015).

\section{Black holes in galaxies}

It is well established now that each galaxy hosts a supermassive black hole
in its center, which mass is about 1-2 10$^{-3}$ of the bulge mass. This proportionality
has been explained by the star formation moderation exerted by the AGN activity.
 Several mechanisms have been invoked for this feedback action, 
in particular radio jets could be efficient enough (Wagner et al. 2012). Recent
observations have revealed the existence of massive molecular outflows
around active nuclei  (Feruglio et al. 2010, Cicone et al. 2014).
The rate of outflows can be as high as 1-5 times the star formation rate,
meaning that the star formation is indeed reduced.

One environment where the AGN feedback is obvious is galaxy clusters,
and in particular cool core clusters. In most of them, a radio-loud AGN
is hosted by the Brightest Cluster Galaxy (BCG), whose jets dig cavities 
in the hot X-ray emitting plasma (Fabian et al. 2003). In the prototypical example,
the Perseus cool core cluster, cold molecular gas has been detected
around the cavities, corresponding to the expected cooling flow
(Salome et al. 2006). Instead of cooling towards the center, as in the simplistic
model, the gas is cooling at 20-30kpc from the BCG center, and falls along filaments,
that are conspicuous in H$\alpha$.  Radio continuum emission gives
important information about filaments and magnetic fields. This work can
be done at much higher redshift with SKA.

Not all black holes have been seen. According to the proportionality relation
between bulges and black hole masses, there should exist intermediate-mass black holes
(IMBH) in dwarf galaxies, or may be in the outer parts of galaxies, since 
they are not massive enough to spiral into their galaxy nucleus.
One candidate of such an object is the ultra-luminous X-ray source
HLX1, discovered as a companion of the edge-on early-type spiral
ESO 243.49, at 95 Mpc distance. The energy of the X-ray source,
of 10$^{42}$ ergs/s is difficult to explain with binary stars, and 
must come from a black hole of mass 10$^2$-10$^5$ M$_\odot$
(Webb et al. 2010). With SKA, the search for IMBH could be 
more successful.

Smolcic et al. (2015) explore the possibilities opened by SKA and its various sky surveys
in detecting radio-loud and radio-quiet AGN as a function of cosmic time. The question of
bimodality between these two categories might find a solution, in the evolution of quasars
between these two phases, or in different emission mechanisms. The black-hole accretion rate,
and its cosmic history will be established on solid grounds, together with the AGN role
on quenching star formation in galaxies.

\section{Pulsars}

Pulsars are rotating neutron stars, discovered by Bell \& Hewish (1968).
They are the life end of massive stars, after their explosion
in supernovae, if their core is not massive enough to form a black hole.
Their size is of the order of 10km, and their mass 1-2 M$_\odot$, so their 
central density is larger than that of atom nuclei, of the order
of 10$^{15}$g/cm$^3$! The gravity at their surface is
$\sim$ 10$^{11}$ g, and their magnetic field up to B=10$^{12}$ Gauss.

Up to now, 2000 'normal' pulsars are known in the Milky Way. Their rotation
period is usually on the scale of one second (the 
Crab pulsar has a period of 0.03s), acquired just after SN explosion.
But there exists a category of milli-second pulsars (MSP),
 which have been re-activated in X-ray binaries.

Alone the pulsar can live 100Myr, losing energy through multipole radiation.
But in a binary, the companion can
transfer mass and angular momentum when in the giant phase,
accelerating the pulsar. Since the magnetic field B is down to 10$^8$G, the spinning can
live during Gyrs.

Pulsars are exceptional clocks, that can be measured with extreme precision.
The discovery of binary pulsars (Hulse \& Taylor 1975) 
and their timing has already indicated the existence of gravitational
waves (Nobel prize in 1993), since the binary loses energy through emission
of these waves. From the high precision clock and its Doppler effect in the 
binary orbit, it has been possible to measure the shrinking of the orbit.

The extreme precision and high sensitivity to be reached by SKA
will enable progress in the 
physics of accreting white dwarfes, neutron stars and black holes, and in 
general the physics of condensed matter
with strong magnetic field. SKA will open and enlarge many domains, as the emission
mechanisms of pulsars, nuclear physics and strong interaction, transient phenomena
as Rotating Radio Transients (RRAT) or Fast Radio Busrts (FRB), strong field 
gravity and large-scale structure (Antoniadis et al. 2015).

\subsection{Timing of pulsars}

The MSP J0437-4715, one of the best measured milli-second pulsar,
has now (15 July 2014) a period of
P= 5.7574518589879ms $\pm$1 in the last digit (13th).
This digit increases by 1 every half hour.
The neutron star is slowing down due to loss of energy by radiation and
emission of a relativistic wind.
The first 6 digits will keep the same for 10$^3$ yrs.
The time of arrival (TOA) of the signal has been measured with micro-second
precision during several years, and that is why a precision of
14 digits is obtained.

Pulsars provide the most precise measures in Astrophysics. For example
radial velocities in a binary can be estimated with precision of mm/s,
better than the 1m/s precision of exoplanet searches. This can be obtained
after one year of astrometric precision on position, and also on spin down,
 and determining the orbit of the binary (excentricity, peri-astron, orbital period).

The discovery of thousands or more pulsars will require a lot of data
(Cordes et al. 2004).
The interstellar medium (ISM) along the line of sight produces
a dispersion of the pulses, which arrive with a time delay $\Delta t$
depending on the frequency $\Delta t \sim \nu^{-2}$.
Thousands of frequency channels will be observed and delayed, over
a 3GHz bandwidth. A discovery implies to sweep over a whole range of dispersions,
processing Petabytes of data. A few 10$^4$ trial of dispersion measures
are required, and a computing power of 0.1 Petaflops. 
In addition, the de-dispersion is very I/O intensive
(Barsdell et al. 2012).

\begin{figure}[tbp] % figures (and tables) should go top or bottom of
                    % the page where they are first cited or in
                    % subsequent pages
\centering
\includegraphics[width=1.0\textwidth]{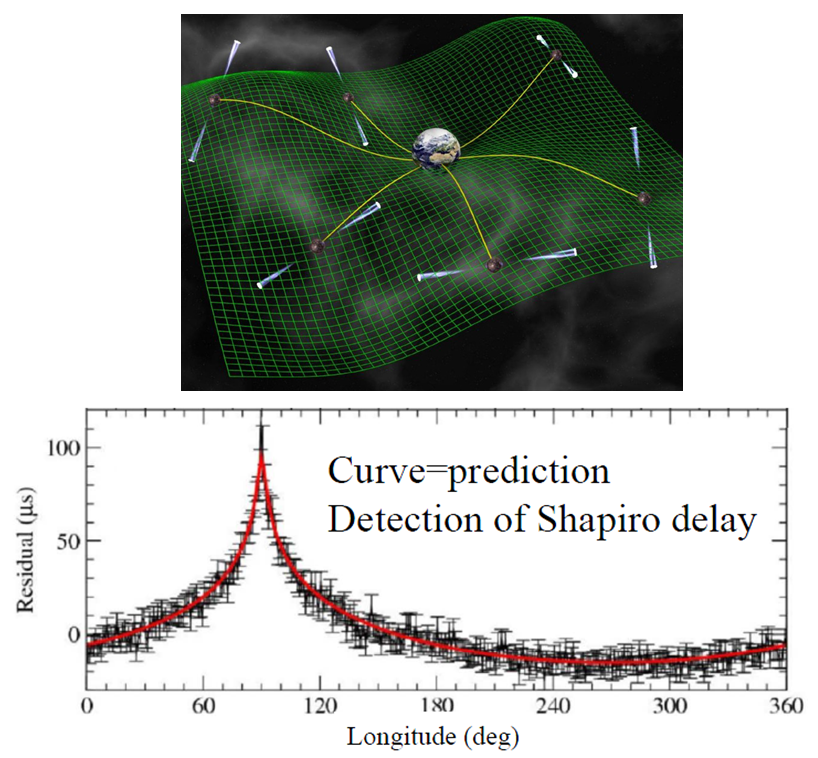}
\caption{{\bf Top} Schematic reception of the pulsar signals
from various regions of the Universe, enabling us to detect 
the space perturbations, and deduce the passage of gravitational waves.
 Image credit: David J. Champion.
 {\bf Bottom}  Measurement of the Shapiro delay on the binary pulsar  PSR J0737-3039A/B,
demonstrating the curvature of space-time.
The difference between observed and predicted arrival times for the pulses are plotted as 
a function of longitude for the pulsar PSR J0737-3039A.  The strong peak at 90$^\circ$
longitude corresponds to the position of the pulsar A behind its companion B, so that the pulse experiences a
delay when moving through the curved space-time near B.
From Kramer et al. (2006).}
\label{fig:pulsars}
\end{figure}

When the binary is edge-on, which is the case of  J1614-2230, there
must exist a 
gravitational delay when the MSP passes behind the white dwarf.
This is called the Shapiro delay: for an orbit of 
8.7 days, the pulses are expected with a delay of 30 $\mu$s!
The delay has been observed with the GBT telescope,
and the GBT-GUPPI instrument, for which
GPU and FPGA are available to process the signal,
 see figure~\ref{fig:pulsars}.

\subsection{Gravitational waves}

 The extreme precision of the timing of pulsars
opens a great opportunity to try to detect gravitational waves,
may be even before the specialised instruments, such as LIGO.
 The principle is to observe and monitor several MSP,
and build an equivalent "instrument", the 
PTA: pulsar timing array (cf figure~\ref{fig:pulsars}, top).
Gravitational waves have nanoHz frequencies  (wavelength of $\sim$ a light-yr).
At these time-scales, the correlation between the time of arrival of the signal
of the array of pulsars will trace space streching.

Gravitational waves coming from the merger of
black holes, if one occurs nearby, would be detectable
at other wavelengths. For the whole of black-hole mergers, 
since they are at more remote distances,
only the  noise due to the ensemble of their
mergers could be detected as a stochastic background.

Already in the recent years, the number of detected MSP has 
increased considerably from 10 in 1995 to more than 200 today.
Many recent ones were discovereed thanks to the Fermi directed radio surveys.
With SKA and precursors, there is a bright future for pulsars discovery.
 The amount of data and beams will be huge, with Terabytes per second.
It might be impossible to record everything, and on the fly processing
is necessary. Instead of re-analysing data, it will be better to re-observe.

\subsection{Tests of general relativity}

Pulsars measured with extreme precision are the occasion to
probe gravity in strong fields. This will be done in binary systems,
either a pulsar and a neutron star, or a black hole.
The standard gravity law can be checked, and also 
the cosmic censorship conjecture, that the black hole is 
a naked singularity, and has no hair, only defined by two quantities, its
mass M and angular momentum J.  The
double pulsars timing will yield 0.05\% test of general
relativity in strong field.

Presently the most
precise data comes from the triple system
PSR J0337+1715 (Ransom et al. 2014). There is in an inner orbit of 1.6 days a pulsar and
a young hot white dwarf of mass 0.2 M$_\odot$, and in an outer
orbit of 327 days, a cool old white dwarf of mass 0.41 M$_\odot$.
 The measurement of all parameters of the system
allows to test the Strong Equivalence Principle (SEP), which appears to
 be verified in strong gravity also, up to now. 
Other scalar-tensor theories have been constrained
(Freire et al. 2012, Antoniadis et al. 2013).

It is expected that 
30 000 normal pulsars exist in the Milky Way, and 20 000 will be
discovered with SKA, together with 10$^4$ MSP
and RRATs  (Rotating Radio Transients). The latter have a more 
irregular cycle but might be more abundant.

\section{SKA: the instrument}

The SKA project, of which a first phase is expected to be
operational in 2020, is the project of a giant radiotelescope 
in the centimetre-metre wavelength range, with 
 one square kilometre collecting surface. This will be
   50-100 times more sensitive than present radio telescopes 
for spectral line observations, and 
 1000  times more sensitive for continuum observations. The
expected range of frequencies is 50 MHz- 25 GHz  (or wavelength 1.2 cm- 6m),
 the field of view from 1 (up to 100) square degrees  at $\lambda$ 21 cm / 1.4 GHz,
with 8 independent fields of view, or beams,
selected within this area, and
 angular resolution up to  0.01 arcsec  at  $\lambda$ 21 cm / 1.4 GHz, obtained
 with baselines up to $\sim$ 3000 km.
The point source sensitivity  will be of
10 nano-Jy in 8 hours of integration.
The optimal antenna configuration is a dense and compact core,
with extended array of antennae disposed in a spiral structure.

At these cm-m wavelengths, the new technology of a wide field of view, 
allowing to observe with 
multi-beams several fields of view simultaneously, is a revolution, 
providing a great efficiency to the observations.
 Beams are formed electronically, with digital delays, for a fixed network
of antennae. The prototype EMBRACE
(Electronic MultiBeam Radio Astronomy ConcEpt) is currently 
tested in Nancay and Westerbork.
The LOFAR telescope is presently using similar
multibeam technology, and serves as a path-finder for SKA. 

SKA is a world-wide project: 55 institutes from 19 countries
are participating, 150 scientists and engineers are involved in the project,
corresponding at present to more than 100 FTE/year on R\&D activities and construction.
The total estimated SKA construction cost is about 1.5 billion euros, while the first phase
SKA-1 is estimated at 650 millions. 

In 2012, two sites have been selected to host the SKA project, Australia and
South Africa. The three radio telescopes being built 
on the two selected SKA sites are called the SKA precursors, they are MWA and
ASKAP in Australia, and MeerKAT in South Africa.  They have vocation to be included
in the final instrument, contrary to the 
SKA Pathfinders. The latter are facilities or instruments that contribute 
R\&D or other knowledge of direct use to the SKA (e.g., LOFAR). 

The foreseen various phases of the different frequency instruments are
listed and defined in the  table~\ref{tab:SKA1}. The expected
time-scales are:
\begin{itemize}
\item  2018 - 2021: construction of SKA1
\item  2019/20: early science begins
\item  2022- 2025: construction of SKA2
\item  SKA operational for 50 years.
\end{itemize}
It is recommended to follow the evolution of the road map
on the web site skatelescope.org.

\begin{table}[tbp]
\caption{SKA Phases 1 and 2$^1$}
\label{tab:SKA1}
\smallskip
\centering
\begin{tabular}{|lccc|}
\hline
             &SKA1-mid    &SKA1-low      & SKA1-survey\\
 where        & Africa    & Australia     & Australia \\
 telescopes   & 254 dishes & 1024 stations  & 96 dishes \\
             & 64 MeerKAT  &  array       & 36 ASKAP \\
             & 190 SKA     &              &  60 SKA \\
with respect to &JVLA/meerKat    &LOFAR   &ASKAP\\
Sensitivity     &   6 xJVLA   &    16xLOFAR   &    6xASKAP\\
Survey Speed    &     74      &     520       &      22 \\
             &SKA2-mid    &SKA2-AA      & SKA2-Low\\
 where        & Africa    & Africa     & Australia \\
 telescopes   & 2500 dishes & Mid-freq aperture array & Low-freq aperture array \\
\hline
\end{tabular}
\\$^1$Based on numbers before the 2015 re-baselining, which resulted in deferring the 
SKA1-survey, reducing SKA1-mid to 70\% and SKA1-low to 50\% of its original size
\end{table}

\subsection{Three SKA precursors}

SKA precursors are instruments currently built on the two sites of SKA, Australia
and South Africa.
Two of the instruments (ASKAP and MeerKAT) are concerned with the frequency range
0.7-1.8 GHz, corresponding to HI-21cm up to z=1, and the third one (MWA) with
the low frequency range (80-300 MHz). 

MWA, the Murchison Widefield Array in Australia, consists of 
2048 dipole antennas arranged as 128 tiles,
each 4x4 array of dipoles. The array is fixed, and the beam is formed by electronics only,
with a field of view of 25 degrees at 150 MHz. Most of the antennas are within 1.5km, yielding
a resolution of a few arcminutes.

ASKAP in Australia is composed of
36 x 12m parabolic antennas, corresponding to a collecting surface of 4000 m$^2$.
They are equipped by multi-beam Phased Array Feeds (PAF), providing a field-of-view of  30 square degrees.
The instantaneous bandwidth is 300 MHz. The instrument is
optimised for 30 arcsec resolution.

MeerKAT in South Africa is composed of
64 x 12m parabolic antennas, corresponding to a collecting surface of 7000 m$^2$.
They are equipped by single-pixel feeds, with a field-of-view of 1 square degree.
The instantaneous bandwidth is 1 GHz.
The instrument is versatile in spatial resolution, between  6 and 80 arcsec.

\begin{figure}[tbp] % figures (and tables) should go top or bottom of
                    % the page where they are first cited or in
                    % subsequent pages
\centering
\includegraphics[width=1.0\textwidth]{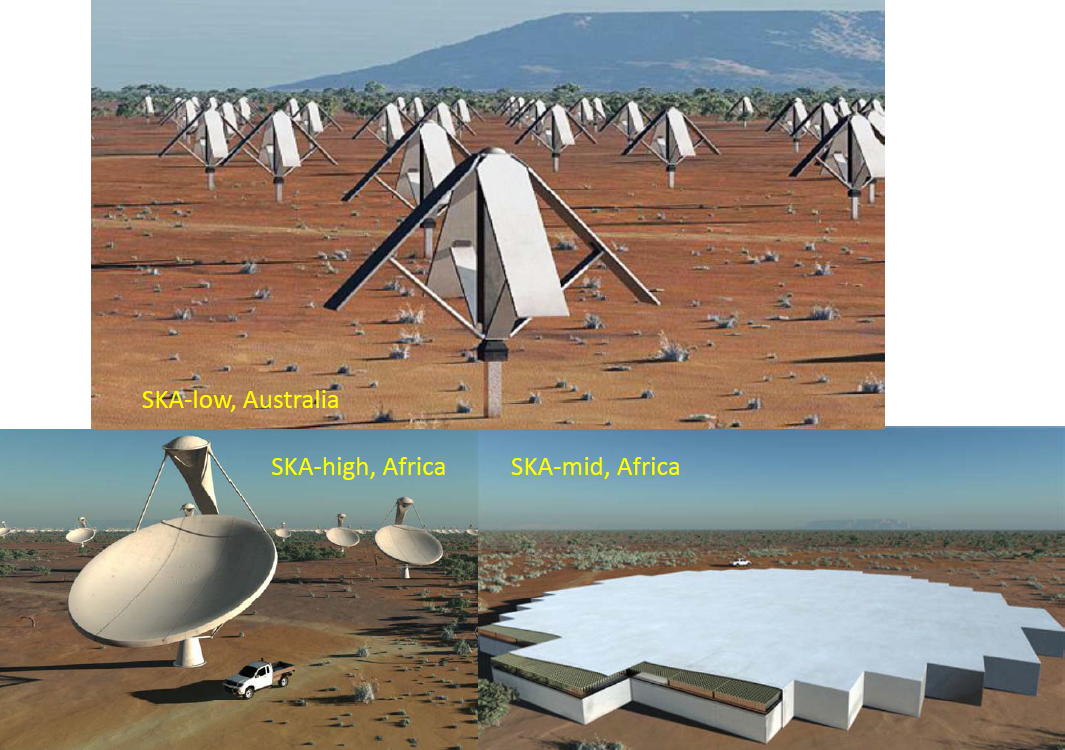}
\caption{{\bf Top} Design of the low-frequency part of the SKA array (to be
built in Australia).
 {\bf Bottom}  Design of the high-frequency and mid-frequency parts,
to be built in Africa.
From skatelescope.org.}
\label{fig:design}
\end{figure}

For all these instruments, the construction has started, 
and is expected to be fully operational in 2016-17.
The selected design can be seen in 
 figure~\ref{fig:design}.

The two higher frequency SKA precursors are very complementary:
ASKAP has a-large field of view, to make all-sky, relatively shallow surveys,
while MeerKAT has a smaller field of view, 
adapted to deeper surveys, at higher or lower spatial resolution.
 These can be compared to present instruments in the northern
hemisphere. In the Netherlands, Westerbork (WSRT) will be implemented
with focal plane arrays (APERTIF), which will increase its field of view
by a factor 25. The overlap in the sky with the southern hemisphere precursors
concerns the declination range +25$^\circ$-30$^\circ$.
The VLA, in the USA, can perform deep integration of small field of
views, down to -40$^\circ$.
The survey speeds at 0.7 and 1.4 GHz of the SKA precursors are already
an order of magnitude larger than that of VLA, and in the SKA-1 phase, they
will be 3 orders of magnitude larger. The SKA-2 phase instruments will be 1-2 orders
of magnitude faster than SKA-1 instruments.

\subsection{Data management}

The high rate of data acquisition is
a huge challenge for SKA. There will be Petabytes per second
to process with Petaflops machines working continuously 
(equivalent to 10$^7$ personnal computers).
The data rate, of the order of exabytes per hour,  obtained with 
the dishes corresponds to 10 times the global internet traffic, 
the data rate with phased arrays is an order of
magnitude larger, and is 100 times the world internet traffic!

In comparison, one of the largest data rates for other instruments
will be encountered with the
LSST (Large Synoptic Survey Telescope), an optical telescope,
which will observe the whole sky every 3 days.  More than half of the cost
is due to the data flow management.  The challenge is however much lighter,
of the order of 20 Terabytes per night, 
leading to a total database of 60 Petabytes after some years of operation.
 It will bring however the new dimension of time in
focussing on the transient sky, publishing 
millions of alerts per night for variable objects (stars, supernovae, etc.).

The SKA will open a new frontier in the Big Data science. The exponential growth of
the amount of data to process will require new methodologies in data intensive astronomy 
(e.g. Taylor 2015).

%\acknowledgments


\begin{thebibliography}{9}
\bibitem{} Abdalla, F.B., Bull, P., Camera, S. et al.: 2015, arXiv:1501.04035, PoS, Advancing Astrophysics with the SKA
\bibitem{} Alcock C., Paczynski B.:  1979  281, 358
\bibitem{} Antoniadis, J., Freire, P. C. C., Wex, N. et al.: 2013,  Science 340, 448
\bibitem{} Antoniadis, J., Guillemot, L., Posenti, A. et al. : 2015, arXiv:1501.05591, PoS, Advancing Astrophysics with the SKA
\bibitem{} Baek, S., Di Matteo, P., Semelin, B. et al.: 2009, A\&A 495, 389
\bibitem{} Barsdell B. R., Bailes, M., Barnes, D. G., Fluke, C. J.: 2012  MNRAS 422, 379
\bibitem{} Bell S. J., Hewish A.: 1969, ApL 4, 211
\bibitem{} Betoule, M., Kessler, R., Guy, J. et al.: 2014, A\&A 568, A22
\bibitem{} Blyth, S-L., van der Hulst, J.M., Verheijen, M.A. et al.: 2015, arXiv:1501.01295, PoS, Advancing Astrophysics with the SKA
\bibitem{} Bouwens, R. J., Illingworth, G. D., Oesch, P. A. et al. : 2011 ApJ 737, 90
\bibitem{} Bouwens, R. J., Illingworth, G. D., Oesch, P. A. et al. : 2015, ApJ, in press
\bibitem{} Brown, M.L., Bacon, D.J., Camera, S. et al.: 2015, arXiv:1501.03828, PoS, Advancing Astrophysics with the SKA
\bibitem{} Burden, A., Percival, W.J., Manera, M. et al.: 2014, MNRAS 445, 3152
\bibitem{} Bull, P., Camera, S., Raccanelli, A. et al.: 2015, arXiv:1501.04088, PoS, Advancing Astrophysics with the SKA
\bibitem{} Chapman, E., Bonaldi, A., Harker, G. et al.: 2015, arXiv:1501.04429, PoS, Advancing Astrophysics with the SKA
\bibitem{} Ciardi, B., Inoue, S., Mack, K. J. et al.:  2015, arXiv:1501.04425, PoS, Advancing Astrophysics with the SKA
\bibitem{} Cicone, C., Maiolino, R., Sturm, E. et al.: 2014 A\&A 562, A21
\bibitem{} Conley A., Guy, J., Sullivan, M. et al.: 2011  ApJS 192, 1
\bibitem{} Cordes J M., Kramer, M., Lazio, T. J. W. et al.: 2004  NewAR 48, 1413
\bibitem{} Einsenstein D. J., Zehavi, I., Hogg, D. W. et al.: 2005  ApJ 633, 560
\bibitem{} Fabian F. A. C., Sanders, J. S., Allen, S. W. et al.: 2003 MNRAS 344, L43
\bibitem{} Feruglio C., Maiolino, R., Piconcelli, E. et al.: 2010, A\&A 518, L155
\bibitem{} Fomalont, E. B., Kellermann, K. I., Richards, E. A. et al.: 1997, ApJ 475, L5 
\bibitem{} Freire, P. C. C., Wex, N.,  Esposito-Far\`ese, G. et al.: 2012 MNRAS 423, 3328
\bibitem{} Geil, P. M., Wyithe, J. S. B.: 2008 MNRAS 386, 1683
\bibitem{} Giovannini, G., Bonafede, A., Brown, S. et al.: 2015, arXiv:1501.01023, PoS, Advancing Astrophysics with the SKA
\bibitem{} Hulse R.A., Taylor J.H.: 1975, ApJ 195, L51
\bibitem{} Iliev, I.T., Mellema, G., Pen, U-L. et al.: 2006, MNRAS 369, 1625
\bibitem{} Jackson J.C.: 2004 JCAP 11, 007
\bibitem{} Jarvis, M.J., Bacon, D., Blake, C. et al.: 2015, arXiv:1501.03825, PoS, Advancing Astrophysics with the SKA
\bibitem{} Kowalski, M., Rubin, D, Aldering, G. et al.: 2008, ApJ 686, 749
\bibitem{} Kramer M., Stairs, I. H., Manchester, R. N. et al.:2006, Science 314, 97
\bibitem{} Leroy A., K., Walter, F., Sandstrom, K. et al.:  2013, AJ  146, 19
\bibitem{} Mellema, G., Koopmans, L., Shukla, H. et a;.: 2015,  arXiv:1501.04203, PoS, Advancing Astrophysics with the SKA
\bibitem{} Obreschkow D., Croton D., de Lucia G. et al. 2009a, ApJ 698, 1467
\bibitem{} Obreschkow D., Kloeckner H-R., Heywood I. et al. 2009b, ApJ 703, 1890 % virtual sky
\bibitem{} Planck collaboration: 2014, A\&A 571, A16
\bibitem{} Pritchard J.,  Loeb A.: 2010: Nature 468, 772
\bibitem{} Ransom, S. M., Stairs, I. H., Archibald, A. M. et al.: 2014, Nature 505, 520
\bibitem{} Salom\'e P., Combes F., Revaz Y.  et al. 2008, A\&A 484, 317
\bibitem{} Smolcic, V., Padovani, P., Delhaize, J. et al.: 2015,  arXiv:1501.04820, PoS, Advancing Astrophysics with the SK
\bibitem{} Sullivan M., Guy, J., Conley, A. et al.: 2011, ApJ 737, 102
\bibitem{} Taylor, A.R.: 2015, IAU Highlights of Astronomy, 16, 677 
\bibitem{} van Weeren, R. J., Roettgering, H. J. A., Intema, H. T. et al.: 2012, A\&A 546, A124 
\bibitem{} Wagner A.Y., Bicknell G. V.: 2011, ApJ 728, 29
\bibitem{} Webb N., A., Barret, D., Godet, O. et al.: 2010   ApJ 712, L107
\end{thebibliography}
\end{document}